\documentclass[showpacs,pra,amsfonts,amsmath,twocolumn,
amssymb,superscriptaddress]
{revtex4-1}
  \usepackage{here}
  \usepackage{subfigure}
\usepackage[]{graphicx, amsfonts, amsmath, amssymb, amstext, latexsym, float,color}
\usepackage{comment}
\usepackage{fullpage}
\usepackage{mathtools}
\usepackage{ulem}
\usepackage{natbib}

\newcommand{\ket}[1]{\left |  #1 \right \rangle}

\newcommand{\eat}[1]{}

\providecommand{\U}[1]{\protect\rule{.1in}{.1in}}

\begin{document}

\title{Role of syndrome information on a one-way quantum repeater using teleportation-based error correction 
}



\author{Ryo Namiki} \affiliation{Institute for Quantum Computing and Department of Physics and Astronomy, University of Waterloo, Waterloo ON, Canada N2L 3G1}\author{Liang Jiang}  \affiliation{Department of Applied Physics, Yale University, New Haven, CT 06511 USA}
 \author{Jungsang Kim}\affiliation{Electrical and Computer Engineering Department, Duke University, Durham, NC 27708, USA}
 \author{Norbert L{\" u}tkenhaus} \affiliation{Institute for Quantum Computing and Department of Physics and Astronomy, University of Waterloo, Waterloo ON, Canada N2L 3G1}\begin{abstract} 
We investigate a quantum repeater scheme for quantum key distribution based on the work by Muralidharan {\it et al., }~\prl~112,~250501~(2014). Our scheme extends that work by making use of error syndrome measurement outcomes available at the repeater stations. We show how to calculate the secret key rates for the case of optimizing the syndrome information, while the known key rate is based on a scenario of  coarse-graining the syndrome information. We show that these key rates can surpass the  Pirandola-Laurenza-Ottaviani-Banchi 
 bound on secret key rates of direct transmission over lossy bosonic channels. 

\end{abstract}

\maketitle

\section{Introduction}


  To explore the possibility of quantum communication schemes over a long distance \cite{Briegel98,Zwerger2015,Azuma2015,rmp-QR11,Razavi09}, an essential  question 
  is whether the secret key generation rate of quantum key distribution (QKD) with  the help of intermediate stations could be better than {\it any} key generation scheme without intermediate stations. A clear criterion for this is to  surpass  the Takeoka-Guha-Wilde (TGW)  bound  \cite{Takeoka2013,Takeoka2014}.
This is an upper bound of the secret key rate per optical mode  over a pure lossy channel, and given by 

  \begin{equation}
\label{TGWbound}
R_{\rm TGW} = \log_2\frac{1+\eta}{1-\eta}, \end{equation}
where $\eta \in (0,1]$ is the transmission of the lossy channel.  
Hence, any key generation over a distance corresponding to the transmission $\eta$ cannot surpass $R_{\rm TGW} $  when there are no intermediate stations. While the TGW bound was suggested to be unachievable, Pirandola,  Laurenza,  Ottaviani, and Banchi (PLOB) have reported  that the corresponding tight bound is given by \cite{Pirandola2015,Namiki2016} 
 \begin{equation}
\label{PLOBbound}
R_{\rm PLOB} = \log_2\frac{1}{1-\eta}. \end{equation}
  It has been shown that the TGW bound cannot be overcome if we are only able to use Gaussian channels as intermediate stations in a one-way structure \cite{Namiki2014}. 
  This no-go statement for Gaussian repeaters holds also for the PLOB bound. 
 An open question is whether there are other simple intermediate stations  facilitating quantum repeater behavior. 

Recently, various quantum repeater architectures have been studied \cite{munro12,Sre14,Sre15,Munro2015}. 
Ultimately, one-way schemes with  a  teleportation-based error correction (TEC) approach \cite{knill2001scheme,knill2005} have an advantage in terms of achievable rates.
Due to the structure of the error correction, the  syndrome information of all intermediate stations is available for optimizing the key rate.  In Ref.~\cite{Sre14},  the syndrome measurement was used to flag success or failure events, and to reduce the effective errors in the success case. The secret key rate was then analyzed by calculating the probability that all intermediate stations show success events and by calculating the expected remaining error rates. No further details of the syndrome measurements have been used.  
Hence,  an attainable key rate is immediately determined without the need for keeping track of every combination of possible syndrome outcomes coming through all intermediate stations. This theoretical simplicity  also suggests a relatively low technical difficulty  for a practical implementation. 
%
On the other hand, such a coarse-grained treatment of the syndrome outcomes will discard some of useful information, and the key rate will be  lower than potentially achievable performance of  a one-way protocol that makes use of the fine-grained syndrome information. 
In other words, 
 we can obtain   
a better key rate  
 when we keep the syndrome information and optimize its use. 
 The question is how significant this improvement is so that one can decide whether it is worthwhile to invest the additional processing overhead required for the fine-grained treatment.  

 In this paper,  we show how to calculate the secret key rate of  one-way schemes when the syndrome measurement outcomes are taken into account, and address the question of whether they can be potentially useful in overcoming 
  the PLOB bound. We positively answer this question by showing that one can beat the PLOB bound by making an appropriate choice of the parameters. 
 We also point out that our intermediate stations are regarded as quantum channels, and there exist simple quantum-channel stations which facilitate the behavior of a quantum repeater. 
  Although a fine-grained treatment of the  syndrome outcomes is  not necessary to surpass the PLOB bound, it turns out to be   useful to extend the transmission distance and improve error tolerance  for a high rate key generation above the PLOB bound.  
 

  %

This paper is organized as follows. After an introduction of the models  and  related basic notions of one-way stations, we show how to calculate the  secret key rate with the use of syndrome information in Sec.~\ref{SecII}.
We compare the key rates for various parameters and identify the area where the performance of the TEC stations surpasses the PLOB bound in Sec.~\ref{SecIII}.  There, we also describe the design of a quantum channel station that is sufficient to surpass the PLOB bound.  We summarize the results in Sec.~\ref{SecIV}. 

\section{Structure of intermediate stations and  secret key rate} \label{SecII}
\begin{figure}[tbhp] \includegraphics[width=0.9\linewidth]{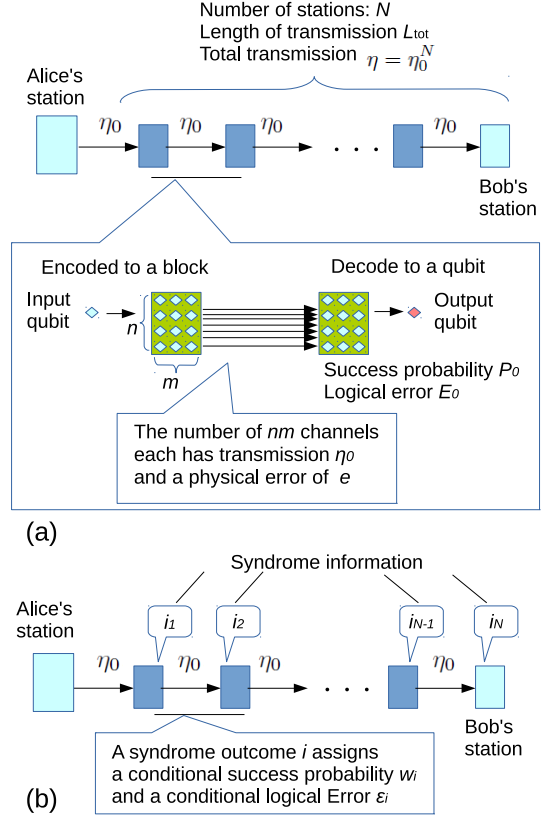}
 \caption{Encoded transmission of a logical qubit based on a block of $N_p=n\ \!  m$ single photons (physical qubits). (a)~Transmission through a concatenation of intermediate stations. Each station is composed of a decoder and an encoder. 
 A unit segment is connected with the number of $n\ \! m$ optical-loss  channels whose transmission is $\eta_0$ and a qubit-error rate is $e$. The efficiency of the unit segment can be described by the success probability $P_0$ the logical qubit is received and the average error $E_0$ of the successfully received logical qubit. (b)~It is not necessary for the intermediate stations to physically execute a decoding operation to output a logical qubit. 
  But, an error correction --- a syndrome measurement followed by a recovery operation--- can be performed  effectively by keeping the syndrome information of intermediate  stations  $\{i_1,i_2, \cdots i_N \} $. A syndrome outcome of a single station  $i$ assigns  conditional success probability $w_i$ and conditional logical error $\epsilon_i$. The set $\{ w_i,\epsilon_i\} $ and  the measured sequence  $\{i_1,i_2, \cdots i_N \} $ determine the net performance of the transmission. 
   }  \label{fig1:fig1.eps}\end{figure}
 
 \subsection{One-way stations and basic mechanism} \label{sec2A}

Let us  consider transmission of a logical qubit using a set of single photon polarization qubits with  repeater stations that perform  quantum error correction 
 as in Fig.~\ref{fig1:fig1.eps}. 
Suppose that the error correction works with a success probability $P_0$ and gives a logical error rate  $E_0$ (averaged over phase and bit) for a lossy segment  of transmission $\eta_0  \in (0,1)$. 
 A concatenation of  $N$  segment-and-station pairs, consisting of loss segments and repeater station units,  has the net success probability 
\begin{align}
&P_{succ} = P_0^N,  \label{nsucc}\end{align} and the net average logical error rate 
\begin{align}
&Q_N= \frac{1-(1-2E_0)^N}{2}. \label{Nerror}
\end{align}  Up to a factor of the protocol efficiency $1/2$, the key rate for the BB84 protocol with this concatenated transmission is given  by  \cite{rmp-QKD09} 
\begin{align}
&K=   P_{succ}\max \left[ \{ 1- 2h(Q_N)\} ,0 \right], \label{r0}
\end{align} where $h(x) = - x \log_2 x  - (1 -x)  \log_2 (1-x) $ is the binary entropy function. The net error of Eq.~\eqref{Nerror} comes from the fact that a link of two binary symmetric channels with the error rates $\epsilon_1$ and $\epsilon_2$ becomes another binary symmetric channel with the error rate
\begin{align}
 G( \epsilon_  {1}, \epsilon_  {2}): = \epsilon_1 (1-\epsilon_2)+ \epsilon_2(1-\epsilon_1).  &  \label{comberr}
\end{align}
 Suppose that the number of the photonic qubits  transmitted per logical qubit  is $N_p$. Since a polarization qubit uses two modes, the number of modes physically  used in this transmission is $2 N_p$. 
 This implies that the key rate per mode is given by 
\begin{align}
&R= \frac{K}{ 2 N_p} .  \label{bpm}
\end{align}

On the other hand, the key rate due to a direct transmission of a single photon over the pure lossy line with the transmission $\eta = \eta _0 ^ N$ is given by 
\begin{align}
&R_d= \frac{1}{2}\max \left[  \eta (1- 2h(0)) ,0 \right]= \frac{1}{2} \eta_0^N, \label{DT1}
\end{align} where the factor $2$ is due to 
the number of optical modes, again. 
For a long distance $\eta \ll 1 $, the PLOB bound  also  gives the key rate  proportional to the overall transmission 
   \begin{equation}
R_{\rm PLOB} = \log_2\frac{1}{1-\eta}\simeq 1.44 \eta = 1.44 \eta_0 ^N .   
 \end{equation}
By comparing $R_{\rm PLOB}$ and  the expression in Eqs.~\eqref{bpm},  the transmission with the intermediate stations beats the PLOB bound if the condition $K > 2 N_p R_{\rm PLOB}$ is satisfied.  With the help of Eqs.~\eqref{nsucc} and \eqref{r0}, this condition can be rewritten  for a long distance as 
\begin{align}
&P_0 >  \eta_0 \left(  \frac{2.88N_p }{1- 2h(Q_N)} \right) ^{1/N} , \label{KeyRelation1}
\end{align} whenever  $1- 2h(Q_N) >0$ holds. 
If the logical error rate is zero, i.e.,  $Q_N =0 $, we have 
\begin{align}
&P_0 >  \eta_0 \left(   {2.88N_p } \right) ^{1/N} . \label{KeyRe}
\end{align}
Since the $N$-th root rapidly converges to one as $N$ becomes larger, we have a simple relation for  $N  \to \infty$:
\begin{align}
&P_0 >  \eta_0  \label{KeyR}. 
\end{align} Therefore, an essential necessary condition to beat the PLOB bound is that the success probability of the station is larger than the transmission of the associated segment. We can interpret the success probability as the success of the transmission, and $P_0$ is regarded as an effective transmission of the channel.  Thereby,  the main role of the intermediate stations is  to boost the effective transmission. 
  Another essential point is that it becomes easier to beat the PLOB bound when more stations are placed along the total transmission line  as seen in Eq.~\eqref{KeyRe}. An important model of the transmission line  that fulfills $Q_N=0$ is the pure bosonic lossy channel. In such a case, an efficient loss error correction to fulfill  Eq.~\eqref{KeyR} is sufficient to beat the PLOB bound.   
  Note that $N$ is associated with the total distance of  the transmission  as  
\begin{align}
&L_\text{tot} = - L_\text{att} \ln \eta = - N L_\text{att} \ln \eta_0. \label{DB}
\end{align} where $L_\text{att}$  is  the attenuation length. In all following numerical results we will use 
\begin{align}
L_\text{att}= 20 {\rm km}.  \label{atelen}
\end{align}
 
 Note also that essentially the same discussion holds  when we replace the PLOB bound in Eq.~\eqref{PLOBbound}
  with the  TGW bound in Eq.~\eqref{TGWbound} as it holds, for $ \eta \ll 1 $, that 
 \begin{equation}
R_{\rm TGW} = \log_2\frac{1+\eta}{1-\eta}\simeq 2.89 \eta =  2.89 \eta_0 ^N .   
\label{tgwwwww} \end{equation}


\subsection{Error model} \label{errormodelmo}
Although the loss in the bosonic channel has  the dominating impact on the quantum repeater performance, there will also be finite errors associated with controlling matter qubits in the intermediate stations \cite{Sre14}. 
As an effective error model, we  assume that all errors are induced  through the channel and the operations in intermediate stations are perfect. 
 For clarity, we assume each physical qubit suffers a physical error $e $ as in Fig.~\ref{fig1:fig1.eps}(a).  
The qubit channel is described by  
\begin{align}
{\cal E}_\text{qubit} (\rho)  = (1-2e) \rho + e Z \rho Z + e X \rho X .
\end{align} 
In our simple error model we assume the error rate to be symmetric in the sense that the qubit bit error is $e_z = e$  and the qubit phase error is also $e_x =  e$. Similar analysis can be performed for a depolarizing channel \cite{Sre14}.

\subsection{Protocol}
\begin{figure*}[tbhp] \includegraphics[width=0.75\linewidth]{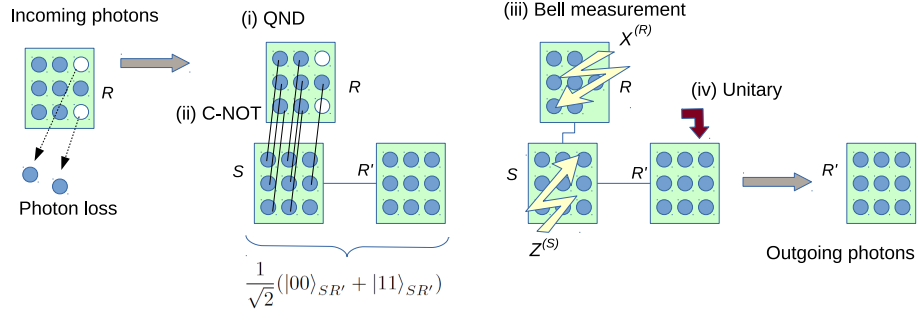}
 \caption{A quantum-teleportation-based error-correction process transfers the logical qubit of incoming photons in the block $R$ into a fresh logical qubit in the block $R^\prime $, and recovers the photon loss. 
 (i)~A quantum non-demolition (QND)  measurement of the photon number is performed to each of the physical qubits in $R$, and the position of the photon loss in the incoming photon block  $R$ is identified. (ii)~A control-not (C-NOT) gate is   applied between  each of the surviving
 photons in  the block $R$ and corresponding photons in the block  $S$ whose logical qubit is prepared to be maximally entangled with the outgoing logical qubit in the block $R^\prime$ as  $(\ket{00}_{SR^ \prime }+\ket{11}_{SR^ \prime })/\sqrt 2$.
     (iii)~A Bell measurement is implemented by logical $X$ and logical $Z$ measurements  performed on the blocks $R$ and $S$, respectively. The Pauli operators of  logical qubits are decomposed into products of Pauli $X$ and Pauli $Z$ of physical qubits. (iv)~Finally, the Pauli frame \cite{knill2005} is adjusted by a unitary operation based on  the Bell measurement outcome. 
     }
  \label{capdfig2.eps}
  \end{figure*}

 
 We consider the one-way scheme~\cite{Sre14} based on a teleportation-based error correction (TEC) \cite{ParityCode04,ParityCide10}.   This scheme is designed to transmit a logical qubit,  such as $\ket{\psi_L} =  \alpha \ket{0_L}  +\beta \ket{1_L} $, encoded in the number of $N_p = n m $ photons with  
\begin{align}
\ket{0_L}&= \frac{1}{\sqrt 2}( \ket{+_L}  +\ket{-_L} )\nonumber \\ 
\ket{1_L}&= \frac{1}{\sqrt 2}( \ket{+_L} -\ket{-_L}  ) \end{align} and 
\begin{align}
 \ket{\pm_L  } 
=  \frac{1}{\sqrt 2} (\underbrace{\ket{00\cdots 0}}_{m} \pm \ket{11\cdots 1} )^{\otimes n}, \end{align}
 where $n,m \ge 2$.  The smallest code uses  four photons, and the unit  photon block of this case is $(n,m) = (2,2)$. We will use the pair ``$(n,m)$''  to specify the block size and thus the code.

Each  intermediate station performs a process of the TEC as in Fig.~\ref{capdfig2.eps}.  First, we perform  a quantum non-demolition (QND) measurement for the incoming qubit block $R$ and identify the positions in the block  where the photon is lost. Second, we apply  control-not (C-NOT) gates between  each of the surviving photons and corresponding photons of the block $S$. The state of the $S$ block is assumed to be prepared  maximally entangled  with the outgoing qubit block$R^\prime$,   $(\ket{0_L0_L}_{SR^ \prime }+\ket{1_L1_L}_{SR^ \prime })/\sqrt 2$ before the C-NOT application.  
Third, we perform a physical $X$ measurement on each photon of the block $R$ and a physical $Z$ measurement on each photon of the block $S$. We obtain the measurement outcomes $X_{i,j}^{R} \in \{\pm 1 \}$ and $Z_{i,j}^{S} \in \{\pm 1 \}$ with $i \in \{1,2, \cdots,n\}$ and $j \in \{1,2, \cdots,m\}$, while  $X_{i,j}=Z_{i,j}=0$ is assigned if the index $(i,j)$ corresponds to the position of lost photons. 
Finally, a unitary operation is performed based on the Bell measurement outcome $(\tilde M_X^R , \tilde  M_Z^S  )$ determined by 
\begin{align}
\tilde M_X^R  &= \textrm{sign} \left[ \sum_{i=1}^n \left( \prod_{j=1}^m X_{i,j}^R\right) \right],  \label{mxmz1}\\
\tilde  M_Z^S  &=  \prod_{i=1}^n  \left[ \textrm{sign}  \left(\sum_{j=1}^m Z_{i,j}^S\right) \right] . \label{mxmz2}
\end{align}
Here, ${\rm sign} (x)$ is associated with the majority vote and assigns $\{-1,0,1\}$ depending on $x <0$, $x=0$, or $x>0$, respectively,  while the product $\Pi$ is associated with the parity.  If either $\tilde  M_X^R = 0$ or $\tilde  M_Z^S= 0$, we say the Bell measurement is inconclusive, and discard the transmission attempt through the quantum repeater chain.

We may classify the statistics of the outgoing qubit based on the pattern of the QND outcomes, namely,  the number of  lost photons and their location. We refer to this information as the pattern component of the syndrome.  Let $S_\text{all}$ be the set of all possible patterns. The performance of the TEC process will be characterized by the syndrome probability $ \{ w_i \}  $  and logical error rates $ \{\epsilon_i \}$ associated with the pattern $i \in S_\text{all}$. 
 To be specific, we are interested in the joint probability that a recoverable pattern appears and the following Bell measurement is conclusive, i.e,  $\tilde  M_{X,Z} \neq 0$, and the logical error rate of such a event.
  We may call the subset of  the patterns being responsible for  such events  the \textit{informative syndromes},  $S_0$.
  A detailed note how to determine this set will be presented in Appendix~\ref{APA}.  
 As we will see in the next section, the key rate over a sequence of stations is  determined by the  observed sequence of patterns. 



\subsection{Key rate} We will sketch how to calculate the key rate  for a sequence of $N$ loss-segment-and-station pairs whose structure is described in Fig.~\ref{fig1:fig1.eps}(b). We assume the BB84 protocol corresponding to  Sec.~\ref{sec2A}.

Let $S_0 = \{1,2,3, \cdots, l\}$ be the set of  informative syndromes of a given intermediate station. 
We denote by  $\{w_i \}_{i \in S_0}$ the set of the success probability, and by $\{ \epsilon_i^{(z)}, \epsilon_i^{(x)}\}_{i \in S_0}$ the associated logical bit error and phase error rates.  
The total measurement probability and the average error rate over the syndromes $S_0$ can be associated with $P_0$  and $E_0$ in Eqs.~\eqref{nsucc}~and~\eqref{Nerror}  as
 \begin{align}
 P_0&=\sum_{i \in S_{0}} w_i , \nonumber \\
 E_0  &=\frac{1}{ P_0}\sum_{i \in S_{0}}w_i  \left( \frac{  \epsilon_i^{(z)}+ \epsilon_i^{(x)} }{2} \right) . \label{e0}\end{align}
In what follows we may refer to the scenario using these averages as the {\it coarse-graining} (CG) scenario that  discards the index$i $ 
 of  informative syndromes. Note that the analysis in Refs.~\cite{Sre14,Sre15} is based on the CG scenario.  On the other hand,  we may refer to the scenario using all indices of  informative syndromes  as the {\it fine-grained} (FG) scenario. 

For a sequence of  two  stations, each transmission of a logical qubit comes with a pair of  outcomes $i_1$ and $i_2 $. This formally specifies the process having  a set  of  a probability   $w_{i_1,i_2}^{(2)}$  and  logical bit-and-phase errors  $\epsilon_  {i_1,i_2 }^{(2,\gamma)}$ with $\gamma \in \{z,x\}$. Here,  ``2'' in the superscript  indicates the number of total stations, and the last station is Bob's station in our notation.  By construction we can write 
\begin{align}
w_{i_1,i_2}^{(2)} &= w_{i_1}w_ {i_2}, \nonumber \\ 
\epsilon_  {i_1,i_2}^{(2,\gamma)}&= \epsilon_  {i_2,i_1}^{(2,\gamma)} = G( \epsilon_  {i_1}^{(\gamma)}, \epsilon_  {i_2}^{(\gamma)}),   
\end{align} 
where   the function $G(\epsilon_  {i_1}, \epsilon_  {i_2})$ is given in Eq.~\eqref{comberr}.

For a sequence of $N$ stations, each transmission of a logical qubit comes with $N$ outcomes $i_1, i_2, \cdots, i_N$ and its statistical property  is formally specified by a probability $w_{i_1,i_2, \cdots, i_N}^{(N)}$ and  logical error rates $\epsilon_  {i_1,i_2, \cdots, i_N}^{(N, \gamma)}$ with $\gamma \in \{z,x \}$. 
  This implies the following expression of the secure key rate
\begin{align}
K=& \sum_{i_1, i_2, \cdots, i_N} w_{i_1,i_2, \cdots, i_N}   \nonumber \\ 
& \times \max \left[0, 1 - \sum_{\gamma \in  z,x }h (\epsilon_  {i_1,i_2, \cdots, i_N}^{(N,\gamma)} ) \right].   \label{KKKK}
\end{align}

For $N$ stations, the number of possible outcomes is $l^N$. Hence, it seems difficult to calculate the key rate for a large $N$ since the number of the relevant terms $\{ (w_{i_1,i_2, \cdots, i_N}^{(N)}, \epsilon_{i_1,i_2, \cdots, i_N}^{(N, z )}, \epsilon_{i_1,i_2, \cdots, i_N}^{(N, x )})\} $ increases exponentially with regard to $N$ as  $2l^N$. 
However, as we can see from  Eq.~\eqref{comberr},  the permutation of indices does not change the value of  $(w^{(N)}, \epsilon^{(N,z)},\epsilon^{(N,x)})$.  This means that  we only need to calculate the triplets  whose indices are  different under the permutation.  By focusing on  the number of the same outcomes we can rewrite the key rate of Eq.~\eqref{KKKK} as 
\begin{align}
K=  \sum_{\sum_ {i=1}^{l} N_i =N  }  w_{\{N_i\}}^{(N)}  \max \left[1 -  \sum_{\gamma \in  z,x } h( \epsilon_{\{N_i \}}^{(N, \gamma)}),0  \right], \label{optratedayo}
\end{align} 
where $N_i$ indicates the number of stations whose measurement outcome is $i$ and 
\begin{align}
w_{\{N_i \}}^{(N)} = &    \frac{N!} {N_1 ! N_2 !    \cdots N_l !} \prod_{i=1}^{l}  w_{i} ^{N_i}, 
\nonumber \\ 
\epsilon_{\{N_i \}}^{(N, \gamma )}  =& G( \cdots ( G ( G (\epsilon_1^{(N_1, \gamma )} , \epsilon_2 ^{(N_2, \gamma )} ), \epsilon_3^{(N_3, \gamma )}),  
\cdots  ,\epsilon_{l} ^{(N_l, \gamma )} )  \nonumber \\ 
=&\frac{1}{2} \left(1 - \prod_{i=1}^l  (1-2\epsilon_{i}^{(\gamma )})^{N_{i}} \right)
, \label{wenet}
\end{align} 
where $\gamma \in \{z,x \}$ and \begin{align}\epsilon_i^{(N_i, \gamma)}= \frac{1-(1-2 \epsilon_i^{(\gamma )})^{N_i}}{2}. \label{Cerror}\end{align}   


Now, we can calculate the key rate for  an arbitrary $N$ sequence of intermediate stations  from  the set of the success probability and logical error rates  $\{  w_{i}, \epsilon_i^{(z)}, \epsilon_i^{(x)} \} $. 
A detailed procedure to determine this set for the case of  the TEC station  with a given set of the physical parameters $(e,\eta_0,N)$ and the code $(n,m)$   is described in  Appendix~\ref{APA}.  In the following numerical calculations, we use a Monte-Carlo method to determine the success probability of the whole sequence of intermediate stations $ w_{\{N_i \}}^{(N)}$  from the set of success probabilities   $ \{ w_{i} \}_{i \in S_0}$.

 

\section{Results} \label{SecIII}
We will calculate the key rate of the TEC stations based on the formula derived in  the previous section and show their potential  as a genuine quantum repeater to provide a higher key rate above  the PLOB bound.

Our main questions are (i) whether or not the one-way protocol due to the TEC stations  can surpass the PLOB bound and (ii) to what extent the syndrome information is significant to  boost the key rate. In addition, if the answers are affirmative, we would ask (iii) how small a code could we use, and whether  the syndrome information helps us to keep the code size smaller?
To this end,  we will make a comparison between the following three different quantities per channel use for a couple of small size codes: (a)~the fine-grained key rate with retaining all syndrome information, which we refer to as the key rate of the FG scenario, $R_{\textrm{FG}}$, (b)~the coarse-grained key rate of the CG scenario~$R_{\textrm{CG}}$, and (c)~the PLOB bound $R_{\textrm{PLOB}}$  in Eq.~\eqref{PLOBbound}.   To be specific, as in  Eq.~\eqref{bpm}, we define the FG rate per mode as   \begin{align}R_{\textrm{FG}} = \frac{K}{ 2nm} \end{align}
 with $K$ as defined in  Eq.~\eqref{optratedayo}. 
Moreover, we define the key rate per mode for the CG scenario
\begin{align}
&R_{\textrm{CG}} =   \frac{P_{0}^N \max \left[ \{ 1- 2h(Q_N)\} ,0 \right] }{2 nm } \label{rFCG}
\end{align}  with $P_0$ and $Q_N$ as defined in Eqs.~\eqref{Nerror}~and~\eqref{e0}. This key rate $R_{\textrm{CG}}$ is the rate averaging all informative syndromes. 

\subsection{The smallest block encoding (2,2)} \label{resultA} 

\begin{figure}[hptb] 
\begin{minipage}[t]{.45\textwidth}
 \includegraphics*[width=  \linewidth]{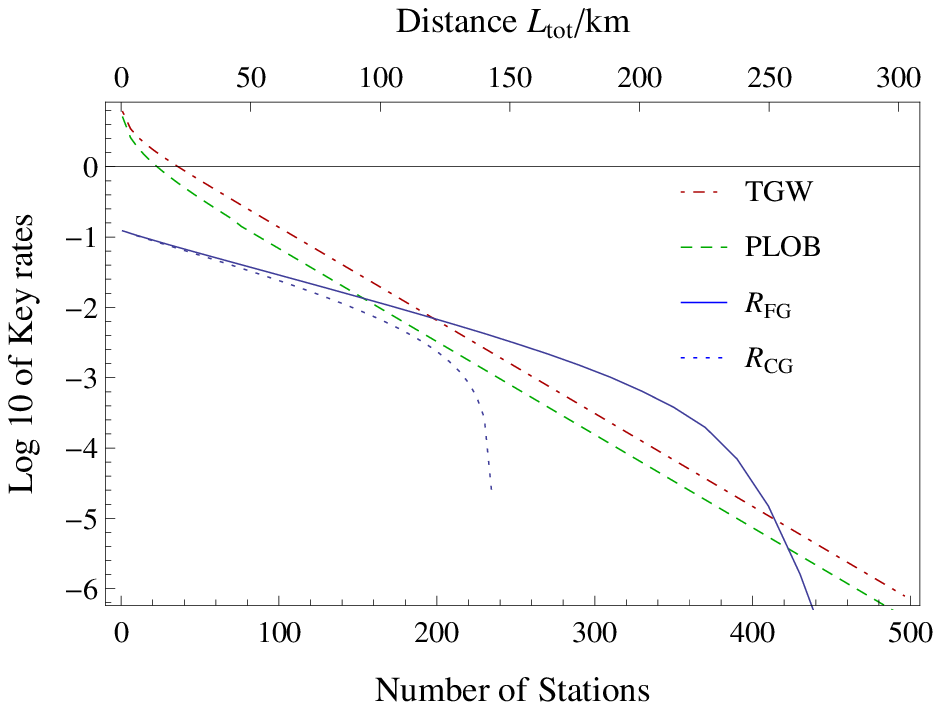}
   \caption{The key rates  $ (R_{\textrm{TGW}},R_{\textrm{PLOB}}, R_{\textrm{FG}}, R_{\textrm{CG}})$  for  $\eta_0= 0.97$ as functions of the number of stations $N$. At $N \simeq 220$, the key rate of the CG scenario $R_{\textrm{CG}}$ drops rapidly  without any crossover to the PLOB bound. 
 For the  FG key rate $R_{\textrm{FG}}$, there is a crossover with the PLOB bound  $R_{\textrm{PLOB}}$.  
 The FG key rate shows relatively slow decay until  $N \simeq 400$. \label{mainG1s1}}
 \end{minipage}
 \hfill
\begin{minipage}[t]{.45\textwidth}
 \includegraphics*[width=  \linewidth]{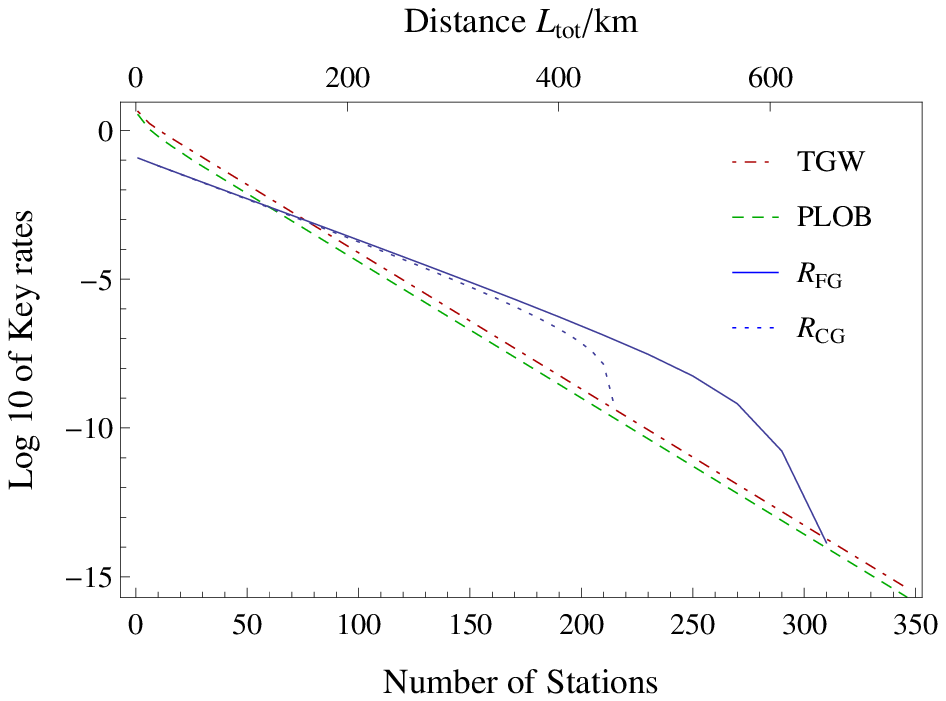}
   \caption{The key rates  $ (R_{\textrm{TGW}},R_{\textrm{PLOB}}, R_{\textrm{FG}}, R_{\textrm{CG}})$  for  $\eta_0= 0.90$ as functions of the number of stations $N$. For $N \simeq 210$ ($L_{\rm tot}\simeq 440$km), the CG key rate  drops rapidly, whereas the FG key rate sustains until $N \simeq 300$ ($L_{\rm tot}\simeq 630$km). 
     \label{mainG1s2}}   \end{minipage}
\end{figure}

\begin{figure}
\begin{minipage}[t]{.45\textwidth}
\includegraphics*[width=  \linewidth]{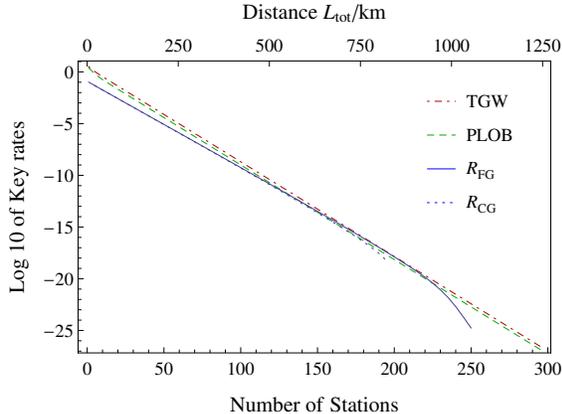}
 \caption{(Color online) The key rates $ (R_{\textrm{TGW}},R_{\textrm{PLOB}}, R_{\textrm{FG}}, R_{\textrm{CG}})$  for  $\eta_0= 0.81$   of the number of stations. The FG key rate $ R_{\textrm{FG}}$ marginally goes up to the PLOB bound $R_{\textrm{PLOB}}$  around  $N \simeq 190$ (which corresponds to $L_{\rm tot} \simeq 800 {\rm km }$), while the CG scenario  could not beat the PLOB bound. \label{mainG1s3}}
 \end{minipage}
\end{figure}

Figures~\ref{mainG1s1},~\ref{mainG1s2},~and~\ref{mainG1s3}  show typical behavior of the key rates for the smallest code $(2,2)$ with the physical error of $e= 5\times 10^{-4}$, where the unit transmission is $\eta_0= 0.97$, $\eta_0= 0.90$ and  $\eta_0= 0.81$, respectively (On the basis of Eqs.~\eqref{DB} and \eqref{atelen}, the unit transmission distance is $L_0 = 0.61\textrm{km} ,  2.1\textrm{km}$, and $4.2\textrm{km} $, respectively). In each of the three graphs, we can observe  regimes  where $R_{\rm{FG}}$ beats  $R_{\rm{PLOB}}$.  In  Fig.~\ref{mainG1s2} ($\eta_0=0.90$)  there is also a regime  where $R_{\rm{CG}}$ beats  $R_{\rm{PLOB}}$ while we could not observe such a regime    in Fig.~\ref{mainG1s1} ($\eta_0=0.97$) and Fig.~\ref{mainG1s3} ($\eta_0=0.81$).  Hence, the use of the syndrome information results in a substantial difference in the parametric regime to beat the PLOB bound. 

These numerical results positively answer  the question (i), namely, the one-way protocol with the TEC intermediate stations beats the PLOB bound in certain parametric regime. Interestingly,  
even the smallest code of intermediate stations can beat the PLOB bound. Moreover, the CG scenario has a lower key rate than the FG scenario by construction, but still has the ability to surpass the PLOB bound. 
These numerical results  also imply that these statements remain essentially the same when we use the TGW bound instead of the PLOB bound.  One can see that the TGW bound stays slightly above the PLOB bound in Figs.~\ref{mainG1s1},~\ref{mainG1s2},~and~\ref{mainG1s3}.

\begin{figure}[H]
\begin{minipage}[p]{.45\textwidth}
\includegraphics*[width= \linewidth]{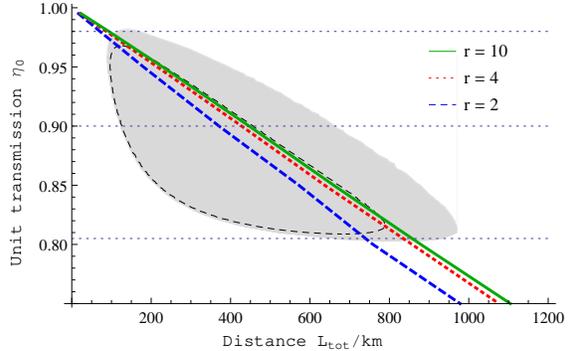}
 \caption{The gray area shows  the regime where the FG key rate surpasses the PLOB bound $R_{\textrm{FG}} \ge  R_{\textrm{PLOB}} $. The dashed area shows  the  CG key rate surpasses the PLOB bound $R_{\textrm{CG}} \ge  R_{\textrm{PLOB}}$.  For the dotted level lines $\eta_0= 0.97$ and  $\eta_0= 0.81$, we can see that the CG key rate  $R_{\textrm{CG}} $ cannot surpasses the PLOB bound, while the FG key rate can do (See  Fig.~\ref{mainG1s1} and Fig.~\ref{mainG1s3}). On the other hand,  the middle of this two level lines shows a  wide regime in which both   $R_{\textrm{FG}}$ and  $R_{\textrm{CG}}$ can surpass the PLOB bound (See the actual behavior of the key rates for the dotted level $\eta_0= 0.90$ in Fig.~\ref{mainG1s2}).  Blue-dashed, red-dotted, and green-solid lines show the boundaries where
  $R_{\textrm{FG}}$ becomes 2, 4, and 10 times larger than 
  $R_{\textrm{CG}}$, respectively,  ($r:= R_{\textrm{FG}}/ R_{\textrm{CG}} $).  The code size is  $(n,m)=(2,2)$ and the physical error rate is $e= 5\times 10^{-4}$.  The areas will shrink when the physical error becomes larger (See Fig.~\ref{mainG2}).   
\label{mainG1}}
 \end{minipage}
\end{figure}

Regarding the second question (ii), we can see from Figs.~\ref{mainG1s1},~\ref{mainG1s2},~and~\ref{mainG1s3} that the scenario utilizing  the syndrome information substantially extends the distance where we can find a positive key. Note that the number of stations $N$ is proportional  to the distance $L_{\rm{tot}}$  through the relations in Eqs.~\eqref{DB}~and~\eqref{atelen}.  Figure~\ref{mainG1} shows an overview of the whole parameter regime  where $R_{\rm{FG}}$ and $R_{\rm{CG}}$ surpass the PLOB bound $R_{\rm{PLOB}}$  in terms of  the total distance $L_{\rm{tot}}$  and the unit transmission $\eta_0$ for $e = 5\times10^{-4}$.   In the shade area,  $R_{\rm{FG}}>R_{\rm{PLOB}}$ is satisfied, while it is inside of the dashed loop that $R_{\rm{CG}}  >  R_{\rm{PLOB}}$ holds.  
The distance where the key rate starts to beat the PLOB bound is not so different between  $R_{\rm{FG}}$ and  $R_{\rm{CG}}$. In contrast, the distance where the key rate falls again below the PLOB bound  for the FG scenario is  substantial longer than that for the CG scenario. Typically,  additionally gained  distance is about $200$km.  
   
 It would be worth noting that this specific example of our scheme beats the PLOB bound in a middle distance, such as $200\textrm{km}  \sim 1000$km,  while the key rate of  a practical single-photon BB84 protocol using threshold detectors  rapidly drops away around $300  $km with a moderate dark count rate \cite{rmp-QKD09}. Therefore, the code $(2,2)$ provides a possible architecture to gain a higher key  rate at such a middle distance  although it requires a substantial number of stations ($\sim 200$), as shown in Figs.~\ref{mainG1s1},~\ref{mainG1s2},~and~\ref{mainG1s3}.  

In order to view the distinctive role of the syndrome information  for gaining a higher key rate over a set of the parameters,  we may
focus on the ratio between the FG key and the CG  key 
\begin{align}
r= \frac{R_{\textrm{FG}}}{R_{\textrm{CG}}}. 
\end{align}
In Fig.~\ref{mainG1}, we also show the lines where this ratio $r$  becomes 2,  4, and 10. 
While a higher value of $r$ does not necessarily mean a substantially higher key rate (because $R_{\textrm{CG}}$ itself may be considerably small), we can ensure a moderate key rate in some cases where we otherwise would not be able to surpass the PLOB bound at all. 
Figure~\ref{mainG1} clearly shows there exists the parameter regime where the following two conditions are simultaneously satisfied: (i) the use of the syndrome information keeps a high  key rate which cannot be achieved by any direct transmission and  (ii) boosts the key rate significantly compared with the key rate of  the CG scenario.  


For a fixed total distance $L_\textrm{tot}$,  we expect a higher number of intermediate stations $N$ would be powerful when there is no physical error. This is because a shorter distance between nearest stations implies a higher success probability. 
If there is a finite physical error ($e >0$), a higher number of stations results in  a higher total logical error  rate because the total logical error accumulates through the action of many stations  (recall the unit transmission is $\eta_0$ and the number of stations is the total distance $L_\text{tot}$ divided by the unit distance $L_0= - L_\text{att} \ln \eta_0$ ). On the other hand, if we separate the intervals further from each other, the loss will increasingly reduce the probability of 
 successful photon detection.   Hence, given the physical error  $e$ and total distance $L_{tot}$, there is an intermediate optimal value of  $\eta_0$ to maximize the key rate. 
In fact, Fig.~\ref{mainG1} shows that neither  too short unit distance nor too long unit distance  gives the key rate better than the PLOB bound. 

\begin{figure}[H]
\begin{minipage}[b]{.45\textwidth}
 \includegraphics*[width=  \linewidth]{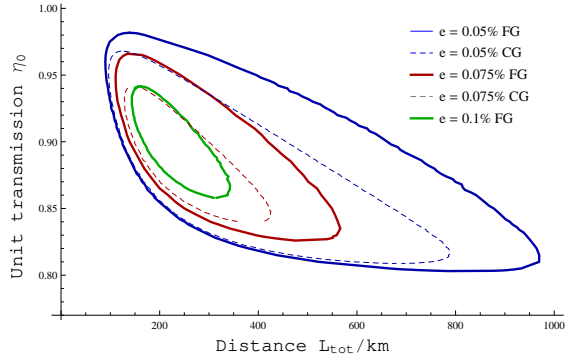}
 \caption{Areas beating the PLOB bound  for the smallest block size $(n,m)=(2,2)$.  The two largest areas for the physical error rate $e = 5.0 \times 10^{-4}$ (the ones in Fig.~\ref{mainG1})   shrink as  $e$ becomes higher. For  $e =1.0 \times 10^{-3}$ in the case of  the CG scenario $R_{\textrm{CG}}$, there is no area where the PLOB bound can be beaten, 
 while  there  still exists a substantial  area to beat the PLOB bound for the FG scenario  $R_{\textrm{FG}}$. 
 The {right} end of the loops for the FG scenario are slightly rugged due to a numerical instability in estimating $R_{\rm{FG}}$. 
  \label{mainG2}} \end{minipage}
\end{figure}

 As the physical error rate $e$  increases, it becomes harder to beat the PLOB bound   
 as illustrated in Fig.~\ref{mainG2}.  The blue-outer solid loop and outer-dashed loop  indicate the regimes $R_{\rm{FG}}> R_{\rm{PLOB}}$  and $R_{\rm{CG}}> R_{\rm{PLOB}}$ respectively, for $e= 5.0 \times 10^{-4}$ (The same areas shown in Fig.~\ref{mainG1}). Both areas shrink for higher physical errors; The shown examples are  the red-inner solid loop and the inner dashed loop for $e= 7.5 \times 10^{-4}$). Moreover,  for $e= 1.0 \times 10^{-3}$, there is no area to beat the PLOB bound for the CG scenario, while we can  observe a small area for the FG scenario. We could not find such an area when $e= 1.5 \times 10^{-3}$ for  the FG scenario. 


\begin{figure}[H]
\begin{minipage}[t]{.45\textwidth}
  \includegraphics*[width=  \linewidth]{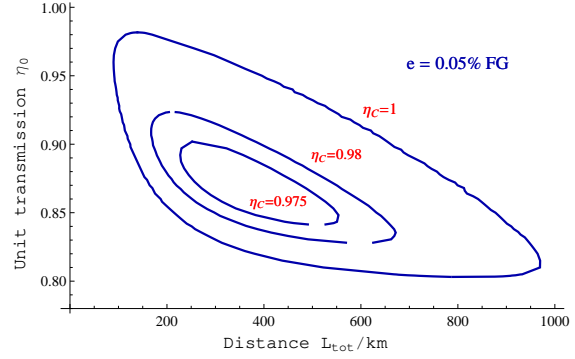} 
\caption{Areas the FG key $R_{\textrm{FG}}$ beating the PLOB bound in the presence of a coupling loss for the smallest code $(n,m)=(2,2)$ with the physical error rate of $e= 5 \times 10^{-4}$.  The Largest loop shows the case without coupling loss ($\eta_c = 1$). The area shrinks for an existence of coupling loss.  
 For a three-percent coupling loss ($\eta_c = 0.97$) we could not find the area.  \label{mainG2Cap98}}
 \end{minipage}
\end{figure}
\begin{figure}[H]
 \begin{minipage}[t]{.45\textwidth}
\includegraphics*[width= \linewidth]{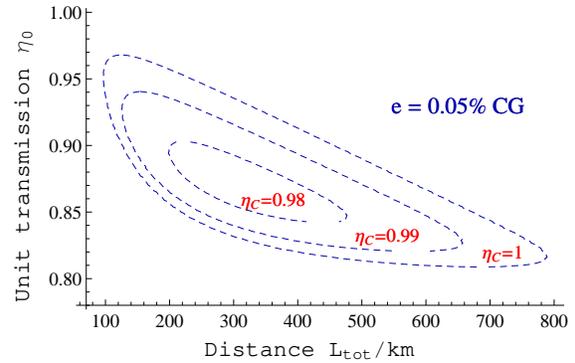}
\caption{The areas  the CG key $R_{\textrm{CG}}$ beating the PLOB bound in the presence of a coupling loss for the smallest code $(n,m)=(2,2)$ with the physical error rate of $e= 5 \times 10^{-4}$.  The largest loop shows the case without coupling loss ($\eta_c = 1$). The area shrinks due to coupling loss. 
  We could not find the area when the coupling loss is  $2.4 \%$ ($\eta_c = 0.976$).
\label{FCG22loss}}
\end{minipage}
\end{figure}

{\textit{Coupling loss--- }}
For practical applications,  it is crucial to estimate the effect of  the coupling  loss.   Suppose that every station exhibits the same coupling efficiency  $\eta_c$ for coupling physical qubits into the channel. 
Then,  the coupling loss will modify the unit transmission  as $\eta_{e\!f\!f} =  \eta_c \eta_0 $ in the calculation of logical errors while we keep  the unit distance $L_0 = -L_{att} \ln \eta_0$ being the same as in Eq.~\eqref{DB}.  In Fig.~\ref{mainG2Cap98}, we show the area surpassing the PLOB bound for $e = 5.0 \times 10^{-4}$ and $(n,m)=(2,2)$ with  the coupling efficiency of $\eta_c = 0.98  $ and $\eta_c = 0.975 $, whereas the largest loop is the lossless case. 
For a three percent coupling loss ( $\eta_c = 0.97$) we could not find such an  area. Similar behavior can be observed for the case of the CG scenario in Fig.~\ref{FCG22loss}. The area vanishes when  $\eta_c = 0.976 $. Consequently, we require a high coupling efficiency $\eta_{e\!f\!f} \sim 99\% $ and a lower physical error rate $e \sim 10^{-4}$ in order that  the one-way scheme with the smallest code $(n,m)=(2,2)$ is experimentally feasible to beat the PLOB bound. 



\subsection{Other small-block codes [(3,2),(3,3),(4,3)]}
\label{resultB}

As we have observed in Fig.~\ref{mainG2}, the physical error rate has to be rather small to beat the PLOB bound via the smallest code~$(2,2)$. More specifically, there are almost no hope to beat the PLOB bound for this block size when the physical error rate is $e \ge 1.5 \times 10 ^{-3}$, even without coupling loss.  Fortunately, we can observe better error tolerance for larger block codes. 

\begin{figure}[H]  
    \begin{center}
 \includegraphics*[width=  \linewidth]{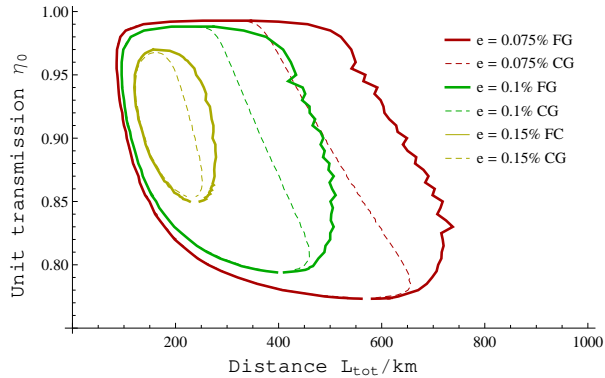} 
    \end{center}
    \caption{Areas beating the PLOB bound for the block size of $(n,m)=(3,2)$ with physical error rates of $e = 7.5 \times 10^{-4}$, $e= 1.0 \times 10^{-3}$, and $e= 1.5 \times 10^{-3}$.  For this block size, we can observe wider areas to beat the PLOB bound compared with Fig.~\ref{mainG2}. 
  In particular, we find  a relatively wide area for the CG scenario with $e = 1.0 \times 10^{-3}$ beating the PLOB bound (the middle green-dashed loop), and there exists such regime even for a higher error rate of $e =1.5 \times  10^{-3}$  (the smallest yellow-dashed loop). The right most boundaries for the FG scenario are  jagged due  a numeric instability. Again, for each of the physical error rates, the FG scenario always extends the longest distance to beat the PLOB bound compared with the CG scenario.  
 \label{mainG3}}
  %
\end{figure}

The second smallest block size is $(n,m)=(3,2)$. Figure~\ref{mainG3} shows the areas that surpass the PLOB bound for a couple of different physical errors $e$.  Compared with the $(2,2)$ code, 
 we can find a substantially wider area to surpass the PLOB bound for both the FG scenario and the CG scenario with  $e =1.0 \times 10 ^{-3}$ (the  green-solid middle loop and the green-dashed middle loop, respectively).
It also gives a finite area for a larger physical error rate of $e = 1.5 \times 10^{-3}$, while there is no such area when the error rate  is  $e = 2.0 \times 10^{-3}$. The shape of areas becomes wider in vertical direction compared with the $(2,2)$ code. This implies that the unit transmission distance between the nearest stations could be larger. Intuitively, the $(3,2)$ code has a better loss tolerance than the $(2,2)$ code. This is because the $(2,2)$ code is unable to maintain the qubit information for loss of two photons while the $(3,2)$ code  still has  certain potential for recovering from two photon loss. Note that the right end of the loops for the FG scenario are rugged due to the numerical instability in estimating $R_{\rm{FG}}$.

\begin{figure}[H]  
  %
    \begin{center}
 \includegraphics*[width= \linewidth]{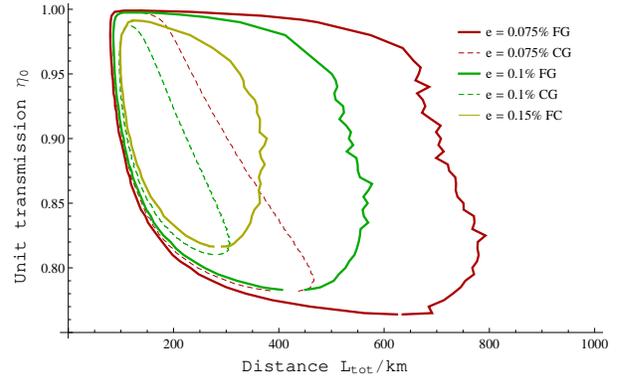}
    \end{center}
  \caption{Areas beating the PLOB bound for the block size of $(n,m)=(3,3)$.   The solid loops  are  for the FG key rate and the dashed loops are for the CG rates. The physical error rates are respectively $e= 7.5 \times 10^{-4}$, $e= 1.0 \times 10^{-3}$,  and $e= 1.5 \times 10^{-3}$  from the largest  solid loop to the smallest solid loop for the FG scenario, while there is no area for the CG scenario with  $e= 1.5 \times 10^{-3}$.  Compared with the block size of $(3,2)$,  the areas for FG key rates grow wider while the areas for  the scenario rather shrink.  The right most boundaries for the FG scenario are  jagged due  the numeric instability. 
 \label{mainG4}}
  \end{figure}

We can observe wider areas for beating the PLOB bound with relatively higher errors by using larger codes. Figure~\ref{mainG4} shows the case of the $(3,3)$ code. In this case, the CG scenario results in smaller areas to beat the PLOB bound when it is compared with the areas for the $(3,2)$ code of Fig.~\ref{mainG3}. Notably, the area vanishes when we use the $(3,3)$ code for  $e = 1.5 \times 10^{-3} $, although the use  of the syndrome information smoothly widens the area.  In general,   the CG scenario unnecessarily smears out the qubit information kept by the measurement outcomes of intermediate stations, and 
thus we find that making the use of fine-grained information can be valuable. Nevertheless, the CG scenario has already shown a significant performance improvement over a long distance  for high coupling efficiency and low error rates \cite{Sre15}. 

\begin{figure}[H]  
  %
    \begin{center}
 \includegraphics*[width= \linewidth]{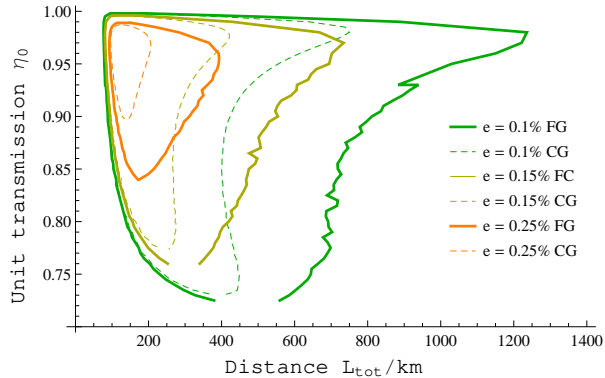}
    \end{center}
  \caption{Areas beating the PLOB bound for the block size of $(n,m)=(4,3)$.  
Solid loops are for the FG scenario whereas the dashed loops are for the CG scenario.   The physical error rates are respectively $e=1.0\times  10^{-3}$, $e= 1.5 \times 10^{-3}$, and $e= 2.5 \times 10^{-3}$, from largest loops to smallest loops.  
  A distinguishing feature is the growth of the areas toward the distance direction around   $\eta_0 \sim 0.97$.  This suggests that one can obtain better performance by using many intermediate stations  with a short unit distance. 
     The right most boundaries are  jagged due the numeric instability. 
 \label{mainG5}}
\end{figure}

The shape of the areas depends on both the block size and the error rate. 
Figure~\ref{mainG5} shows the area to beat the PLOB bound for the case of the  $(4,3)$ code.  An interesting feature observed in this code is that
the area at a higher unit transmission grows significantly towards  
longer distance, and covers relatively wide range of $\eta_0$ and $L_{\text{tot}}$,  although
it still require an order of $10^{-3}$ physical error rate. 
In the case of  $e= 1.0 \times 10^{-3}$, 
we observe a relatively slow decay  of  the upper end of the loop  along  with the line $\eta_0 =1$, which  sustains over the distance of $1000$km.  As we have already mentioned in Sec.~\ref{sec2A}, a better signal transmission  over a long distance can be achieved   when there are no channel errors  by using many stations with  short unit distances. This is because the loss errors can be  corrected well if the distance between the stations is short. 
 The growth and slow decay  along the line $\eta_0=1$ here is regarded  as a concrete example of such an advantage invoking  the PLOB bound.
The slow decay has also been observed for the codes $(3,2)$ and $(3,3)$, although it does not  reach  the long distances over $1000$km (There, the longest distance to surpass the PLOB bound is  achieved with rather lower unit transmission, $\eta_0 \sim 0.85$). Therefore,  the graphs in Figs.~\ref{mainG3}, \ref{mainG4}, and \ref{mainG5} signify the main advantage of using intermediate stations for a high rate secret key generation  with  relatively small block-size codes.


\subsection{Interpretation as  effective quantum channels}
 The performance of the CG scenario can be obtained by a modification of the procedure of the intermediate stations  such that they do not announce the success or fail but send vacuum states for the case of inconclusive events.  With this modification the stations act as quantum channels.  This constructively proves that  there exists  a quantum channel (completely positive trace-preserving map) which works as a repeater station, namely, a sequence of  intermediate stations which works without transmission of classical information can beat the TGW bound as well as
  the PLOB bound. This is in a sharp contrast to the no-go result for the  Gaussian-channel stations \cite{Namiki2014}. 

\section{Summary}\label{SecIV}
We have analyzed potential protocols to utilize TEC one-way quantum repeater stations to increase the key rate over lossy channels beyond the fundamental bounds as given by the TGW bound and the PLOB bound.
We have shown how to calculate the secret key rate of one-way intermediate stations when the syndrome
information of each station is fully available.  
As a general benchmark for potential quantum repeaters, we have compared the performance of the one-way scheme with the PLOB bound, and  numerically identified the parametric area in which our scheme surpasses the PLOB bound for a couple of small block-size codes.  
We have observed that even the smallest block-size code (n,m) = (2,2) with the CG scenario 
 enables us to beat the PLOB bound in a middle distance, although one needs low physical errors $e \sim 10 ^{-4}$ and small coupling loss such as $1\%$. The number of the intermediate stations is typically a few hundred. In our observation, the use of syndrome information helps us to beat the PLOB bound for longer distances, and enables obtaining substantially higher secret key rate.  Our results suggest that the error tolerance will be improved by using  larger blocks. For instance, it has been shown that there is a substantially broad area to surpass the PLOB bound with the $(4,3)$ code for  physical errors around $e \sim 10 ^{-3}$. In this block size,    we have observed  that the distance attained by making use of the syndrome information is significantly longer than the CG scenario. 
In turn,  the performance of  the CG scenario has constructively proven an existence of a quantum channel that works as a quantum repeater. In our construction with the $(2,2)$ code, the quantum channel acts on four photonic qubits, and is  regarded as an eight-mode quantum channel, while it was known \cite{Namiki2014}  that Gaussian channels cannot work as  quantum repeaters even we are allowed to use arbitrary many modes.  It remains open whether our construction is the smallest possible one-way channel stations with respect to the number of modes.
 Overall, we have observed several different aspects to overview the potential performance for QKD with one-way intermediate stations by delving into a couple of small block-size codes with a set of specific  physical error rates. We believe our results are helpful to design a useful architecture for QKD with one-way intermediate stations  in a long run.

\acknowledgments
This work was supported by the DARPA Quiness program under prime Contract No. W31P4Q-12-1-0017, the NSERC Discovery Program, and Industry Canada.

\appendix
\section{Success probabilities and error rates of a TEC station}\label{APA}
In this appendix,  given a code of the TEC stations $(n,m)$ and the error model of Sec.~\ref{errormodelmo},  we show how to determine the set of the success probability and logical error rates  for a TEC station described in Fig.~\ref{capdfig2.eps}. See Ref.~\cite{Sre14} further detail of the TEC stations.  
\subsubsection{Classification of photon-loss patterns}
We consider a unit of $n \times m $ block of photonic qubits. We may call sub-block for a row,  which has $m$ qubits, hence, the total number of sub-blocks is $n$.    We may also use the notation $(n,m)$  to specify the block size.  In what follows, suppose that a block size $(n,m)$ with $n,m \ge 2$ is given.

Due to the transmission loss, each qubit arrives at the next station with the probability of $\eta_0$. Let $n_{LP}$ denote the number of lost photons. For a block of $n \ \!m $ photons, the probability that $n_\text{arri} = n \ \!m -n_{LP}$ photons arrives at the station can be written as
\begin{align}
p_{n_{LP}}:= & \mathrm{Prob} \left[ n_\text{arri} =  n \ \!m  -n_{LP}   \right] \nonumber \\
 = & \binom{n \ \!m}{n_{LP} }\eta_0^{n m -{n_{LP} } }(1- \eta_0)^{n_{LP} }. \label{pnle1} 
\end{align}

\begin{figure}[H] 
 \includegraphics*[width=0.8\linewidth]{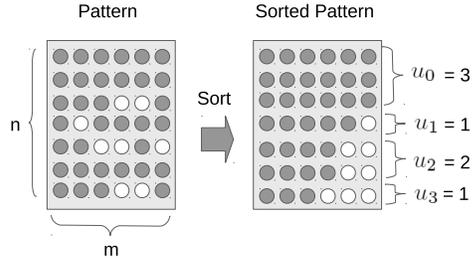} 
\caption{Classification of patterns. The white circles indicate the position of  lost photon. 
For this example, the block size is $(n,m)=(7,6)$ and the classifying index is determined to be  $\vec u= (u_0,u_1,u_2,u_3, u_4, u_5,u_6)^T = (3,1,2,1,0,0,0)^T$. This implies $n_{LP}=\sum_{k=0}^{m-1} ku_k = 8$.\label{patfig1}}
\end{figure}

All patterns of arrived photonic qubits can be classified by a vector $\vec u:= (u_0, u_1, \cdots , u_{m-1})^T$ where $u_k$ represents the number of sub-blocks (rows) with $(m-k)$-arrived photons (See Fig.~\ref{patfig1}). We will sort a given pattern into a sorted form, by moving the position of lost photons right side within each of sub-blocks and shifting sub-blocks that have larger number of photons upward.  Then, from the sorted form we can determine the vector $\vec u$ as in Fig.~\ref{patfig1}. 
  It has to fulfill $\sum_{k=0}^{m-1 } u_{k} = n $ 
 and  $u_0 \ge 1$  in order to satisfy  (i) at least one qubit must arrive for each sub-block and  (ii) at least one sub-block must arrive with no loss, respectively. This condition also implies an acceptable number of lost photons is 
\begin{align}
n_{LP} \le (n-1) (m-1),
\end{align}
where the number of  lost photons  can be expressed in terms of $\vec u $ as 
\begin{align}
n_{LP} (\vec u) = \sum_{k=0}^{m-1} k u_k.  \label{pnle2}
\end{align} 
We refer to the patterns that fulfill  the conditions (i) and (ii) as ``acceptable'' or ``informative'' patterns. For the informative patterns, one can basically correct loss errors and recover the logical qubit in the absence of other types of errors \cite{Sre14}. In practice, there are inconclusive events  that the slot is discarded although the pattern is acceptable. We define the rate that a sorted pattern corresponding to $\vec u$ results in a conclusive event
$q_{\text{conc}| \vec u} $. We set $q_{\text{conc}| \vec u}=0$ if  $\vec u$ cannot be assigned from acceptable patterns.  Complete form of this rate is determined later in Eq.~\eqref{pu}.    

Given an expression of the vector $\vec u$, any acceptable pattern can be  generated from the sorted form by combining (i) permutation of sub-blocks and (ii) permutation of the photon locations within each of sub-blocks.  
Hence,  the number of the possible patterns that gives  a specific form of $\vec u $ 
can be written as 
\begin{align}
\Omega [\vec u] := &\binom {n} {u_0,u_1, u_2, \cdots,  u_{m-1}  }  \prod_{s=1 }^{m-1} \binom {m} {s}^{u_s}   \nonumber \\
=&\binom {n} {u_0}  \prod_{s=1 }^{m-1}  \binom{n-\sum _{k=1}^s u_{k-1}}{u_{s}} \binom {m} {s}^{u_s}. 
\end{align}  


Given the number of photon loss $n_{LP}$, the patterns associated with $\vec u$ appear with the probability of  
\begin{align}
P_{\vec u| n_{LP}} = 
\frac{\Omega [\vec u]}{\binom{nm}{n_{LP}}} . \label{Probu}
\end{align}  
Let $E_{Z| \vec u}$ and $E_{X| \vec u}$ be logical bit error rate and phase error rate associated with $\vec u$, respectively. 
The key rate for a single station can be calculated as 
\begin{align}
&\sum_{\vec u} p_{n_{LP}} q_{\text{conc}| \vec u} P_{\vec u| n_{LP}}  
    \max \left[ 1-  h(E_{Z| \vec u}) -h (E_{X| \vec u}), 0\right]  \nonumber \\
=& \sum_{\vec u}  q_{\text{conc}| \vec u}     \Omega [\vec u] {\eta_0}^{nm -{n_{LP} } }(1- \eta_0)^{n_{LP} } \nonumber \\
& \times   
  \max\left[1-  h(E_{Z| \vec u}) -h (E_{X| \vec u}) , 0\right]  , \label{LQ}
\end{align} where we use Eqs.~\eqref{pnle1}~and~\eqref{Probu}, and $n_{LP}$ is given as a function of $\vec u$ as in Eq.~\eqref{pnle2}. 
In the following part of this appendix, we show how to determine  $q_{\text{conc}| \vec u} $ 
, $E_{Z| \vec u}$, and $E_{X| \vec u}$.

\subsubsection{The logical measurement outcomes}
In the TEC process, we perform individual $X$-measurement whose  outcomes denoted by  $X_{i,j}^R$ for the incoming block $R$ and individual $Z$-measurement  whose outcomes denoted by  $Z_{i,j}^S$ for the local block $S$ (See Fig.~\ref{capdfig2.eps}). We set $X_{i,j}=0$ and $Z_{i,j}=0$ if no photon is detected at the position $(i,j)$ in the QND measurement.  The logical measurement outcomes are determined by
\begin{align}
\tilde M_X^R  &= \textrm{sign} \left[ \sum_{i=1}^n \left( \prod_{j=1}^m X_{i,j}^R\right) \right],  \label{mxmz1}\\
\tilde  M_Z^S  &=  \prod_{i=1}^n  \left[ \textrm{sign}  \left(\sum_{j=1}^m Z_{i,j}^S\right) \right] . \label{mxmz2}
\end{align}
Here, $\textrm{sign} (x)$ is associated with the majority vote, and assigns $\{-1,0,1\}$ depending on $x <0$, $x=0$, or $x>0$, respectively,  while the product $\Pi$ is associated with the parity.  If  $\tilde  M = 0$, we discard the slot. 

\subsubsection{Logical $X$ errors}\label{Apppc}
According to our model in Sec.~\ref{errormodelmo},  each qubit of the $R$ block suffers the phase error $e_x$. Hence, $X_{i,j}^R$ flips with the  probability $e_x$. If all $m$ qubits of $i$-th sub-block arrive, the parity of $i$-th sub-block  $ \prod_{j=1}^m X_{i,j}^R $ is faithfully  observed with the  probability of 
\begin{align}
f_{X0}= \sum_{k=0}^{\lfloor \frac{m}{2}\rfloor} \binom{m}{2k}(1-e_x)^{m-2k } e_x  ^{2k} , 
\end{align} where $\lfloor \cdot \rfloor$ stands for the floor function.

Let $n_M$ be the  number  of  sub-blocks whose photons  all arrived. (Only those blocks join the majority vote.) The probability that the majority vote faithfully gives the logical $X$ is   
\begin{align}
F_{X |\vec u}= \sum_{k=0}^{\lfloor \frac{n_M -1 }{2}\rfloor} \binom{n_M }{k}f_{X0}^{n_M- k }(1- f_{X0})^k   .  \label{27ka}
\end{align}Recall that the number of full sub-blocks is specified by the first element $u_0$ of the vector $\vec u$. 
The majority vote will result in a  draw with the probability of 
\begin{align}
D_{X |\vec u} = \begin{cases}  \binom{n_M   }{n_M  /2  } [e_x ( 1-e_x)]^{n_M  /2 },  & n_M \textrm{ is even}        \\ 0, &  n_M \textrm{ is odd}. \end{cases} 
\end{align} Since we discard the draw events,  the success probability changes by the factor of
\begin{align}
P_{\text{conc}, X | \vec u}:=  {1-D_{X |\vec u}} . \label{pex}
\end{align}  
 After discarding the draw results, the fidelity of the logical $X$ is given by 
\begin{align}
F_{{\rm tot}X |\vec u} = \frac{F_{X| \vec u}}{1-D_{X |\vec u} }.
\end{align}  
This implies the logical $X$-error 
\begin{align}
E_{X| \vec u}= 1- F_{{\rm tot}X |\vec u}= \frac{1- D_{X |\vec u} - F_{X |\vec u}  }{1-D_{X |\vec u} }. \label{xxxx}
\end{align}

\subsubsection{Logical $Z$ errors} \label{Apppd}
According to our model in Sec.~\ref{errormodelmo},  each qubit of the $R$ block suffers the bit error $e_z$. This implies that  qubits in the position of the arrived photons of the $S$ block  suffer the same bit errors $e_z$ due to the action of the C-NOT~gates. 
In order to determine logical~$Z$ we first execute a set of majority votes and then determine logical~$Z$ by taking the parity.

If the number of arrived photons of  the $i$-th sub-block is $m_{M,i}$, the majority vote of this sub-block is faithfully obtained with the probability of  
\begin{align}
f_{Z,i |\vec u}= \sum_{k=0}^{\lfloor \frac{m_{M,i} -1 }{2}\rfloor} \binom{m_{M,i} }{k}(1-e_z)^{m_{M,i}- k }e_z^k    . 
\end{align}
For each sub-block, we can write the probability that  the majority vote is  a draw  
\begin{align}
d_{Z, i |\vec u} =  \begin{cases}  \binom{m_{M,i}    }{m_{M,i}   /2  } [e_z ( 1-e_z)]^{m_{M,i}  /2 },  & m_{M,i} \textrm{ is even}        \\ 0, &  m_{M,i} \textrm{ is odd} \end{cases}. 
\end{align}
Similarly to Eq.~\eqref{27ka} let us define 
\begin{align}
F_{Z, i |\vec u} = \frac{f_{Z,i|\vec u} }{1- d_{Z,i|\vec u} }.
\end{align}
Since we discard the draw events  the success probability changes based on the set of factors:
\begin{align}
P_{\text{conc}, Z,i | \vec u}:=  {1-d_{Z,i |\vec u} }, \label{pez}
\end{align} where $i \in \{1,2, \cdots, n\}$.

Since we take the parity of $n$ sub-blocks, the fidelity of the logical $Z$
 after discarding the draw results is given by 
\begin{align}
F_{{\rm tot}Z|\vec u} = & \sum_{k=0 }^{\lfloor \frac{ n }{2}\rfloor}  \sum_{j=1 }^{ \binom{n}{2k}}\left[ \prod_{i\in s_j^{2k}} (1- F_{Z, i|\vec u} )   \prod_{i \in \bar s_j^{2k}}   F_{Z, i|\vec u}     \right],  
\end{align}
where  $s^{(2k)} = \{s_j^{(2k)}\}_{j=1,2, \cdots, \binom{n}{2k} }$ denotes  the set of all possible subsets of  $s_0= \{1,2,3,\cdots, n-1,n\}$  with the number of elements $2k$ (the length of  each $s_j^{(2k)} $ is   $2k$, and   $j$ runs from $1$ to $ \binom{n}{2k} $). We also define $\bar s_j^{(2k)} =  s_0 \setminus  s_j^{(2k)} $.  
Then, 
the logical $Z$-error is given by 
\begin{align}
E_{Z|  \vec u } = 1-F_{{\rm tot}Z |\vec u}.  \label{zzzz}
\end{align}

\subsubsection{Rate for conclusive events}
Using Eqs.~\eqref{pex} and \eqref{pez}, for any acceptable pattern associated with $\vec u$,  we can assign the pair of  the logical bits $(X,Z)$  with the rate  
\begin{align}
q_{\text{conc}| \vec u} =&   P_{\text{conc},  X  |\vec  u}  \prod_{i=1}^n      P_{\text{conc},  Z,i | \vec u} ,  \label{pu} 
\end{align} and we set $q_{\text{conc}| \vec u}=0$ if $\vec u$ is not associated with an acceptable pattern. 

Now, the set of the success probability and logical error rates  $\{  w_{i}, \epsilon_i^{(z)}, \epsilon_i^{(x)} \} $ for our TEC stations is given by relabeling  the index $i \to \vec u$ as  
 \begin{align}
w_{\vec u} &=p_{n_{LP}} q_{\text{conc}| \vec u}  P_{\vec u | n_{LP}},
\nonumber \\ 
\epsilon_{\vec u}^{(z)}  &= E _{Z| \vec u }, \nonumber \\ 
\epsilon_{\vec u}^{(z)}  &= E _{X| \vec u }.  \label{wesingle}
\end{align}
From this triplet, we can determine the key rate for  a sequence of  intermediate stations  by using  Eqs.~\eqref{optratedayo},~\eqref{wenet},~and~\eqref{Cerror}.


%

\end{document}